\documentclass[twocolumn,twoside]{revtex4}
\usepackage{graphicx}
\usepackage{amsmath}
\usepackage{fancyhdr}
\newcommand{\OpL}[2][{}]{Q_{#2}^{#1}}
\newcommand{\opL}[3][{}]{[O_{#2}^{#1}]_{#3}}

\newcommand{\be}{\begin{equation}}
\newcommand{\ee}{\end{equation}}

\newcommand{\OpLt}[2][{}]{\widetilde{Q}_{#2}^{#1}}
\newcommand{\alS}{\alpha_s}

\begin{document}

\title{Simple Rules for Evanescent Operators in One-Loop Basis Transformations}

%

\author{Jacky~Kumar}
\affiliation{TUM Institute for Advanced Study, Lichtenbergstr. 2a, D-85747 Garching, Germany}

\begin{abstract}
The basis transformations of the effective operators often involve Fierz and other relations which are only valid in $D=4$ space-time dimensions. 
In general, in $D$ space-time dimensions, however, the evanescent operators have to be introduced to preserve such identities. 
Such operators contribute to one-loop basis transformations as well as to two-loop renormalization group running. In this talk, I discussed 
a simple procedure for systematically changing of a basis at the 1-loop level including shifts due to evanescent operators. As an example, we apply this 
method to derive the 1-loop basis transformation from the BMU basis useful for NLO QCD calculations, to the JMS basis used in the matching to the SMEFT.
\end{abstract}

\maketitle

\thispagestyle{fancy}

%
\section{Introduction}
\begin{figure}
\begin{center} 
\includegraphics[width=7cm, height=8cm]{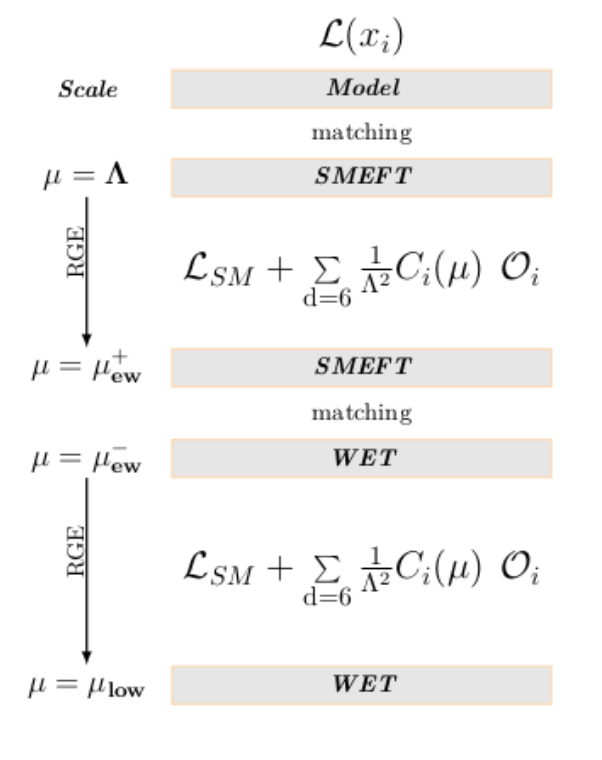}  \label{fig:flow}
\caption{The RGE running in SMEFT and WET.}
\end{center}
\end{figure}

The Standard Model Effective Field Theory (SMEFT) provides a convenient 
framework to parameterize the new physics (NP) effects \cite{Buchmuller:1985jz}. The SMEFT operators are constructed 
using the fields of the SM and are invariant under the full gauge symmetry of the same. 
Further, it is assumed that the electroweak (EW) symmetry in SMEFT is broken using one Higgs-doublet as 
in the SM. In general, the SMEFT Lagrangian can be written as 
\begin{equation}
\mathcal{L}_{\rm SMEFT} = \mathcal{L}_{\rm SM} + \mathcal{L}_{\rm eff}\,,
\end{equation}
here,
\begin{equation}
\mathcal{L}_{\rm eff} = \sum_{\rm d =5,6} C_i O_i. 
\end{equation}
At the dimension-six level there are in total 59 operators in the Warsaw basis \cite{Grzadkowski:2010es} without including different 
flavour permutations. At the NP scale, a renormalizable model generates a unique set of SMEFT 
operators after integrating out the heavy degrees of freedom. 
Then to compute the low-energy observables in terms of these operators, one have to run down the corresponding coefficients (WCs) 
using the RGEs. The running between the NP scale and the EW 
scale is known at 1-loop level \cite{Alonso:2013hga}  \footnote{See also \cite{Aebischer:2022anv} for partial running at the NLO level.}. 
On the other hand in the Weak Effective Field Theory (WET) i.e. the RGE running below the EW scale is also known at the 
NLO level \cite{Buras:2000if}. At the EW scale, the two EFTs need to be 
matched choosing certain bases of operators for each. The complete SMEFT to WET matching up-to 1-loop level
 is known \cite{Jenkins:2017jig,Dekens:2019ept}. The whole procedure is depicted in Fig.~\ref{fig:flow}.
The SMEFT to WET matching has been performed in the JMS basis for WET. However, the QCD NLO RGE running has been computed 
in so-called the BMU basis. Therefore, in order to consistently use these two results a basis transformation 
between the JMS and BMU basis is required. We have presented a simple method to perform such transformation at 
the 1-loop level. As an example, we have provided the complete basis transformation rules between the JMS and the BMU basis at 1-loop 
level in QCD.  More  details can be found in the Ref.~\cite{Aebischer:2022tvz}.
\section{The method}
In this section, we describe the method for the basis transformation at 1-loop level. First, the tree-level 
transformation can be written as 
\begin{equation}
O_{JMS} = \hat  R^{(0)} Q_{BMU}\,,
\end{equation}
here,
\begin{eqnarray}
Q_{BMU} = \{Q_1, Q_2, Q_3, ..., Q_N   \}\,, \\
O_{JMS} = \{O_1, O_2, O_3, ..., O_N   \}\,,
\end{eqnarray}
represent two basis vectors and $\hat  R^{(0)}$ is the tree-level transformation matrix. 
Such basis transformation often require Fierz identities which are valid only in $D=4$ space-time
 dimensions. 
However, at the 1-loop level the divergent current-current and penguin insertions have to be regularized using
the dimensional regularization in $D\ne 4$ dimensions. As a result, the evanescent operators (EVOs)
has to be introduced. The EVOs are defined using the relation
\begin{equation}
{\rm EV_i} = Q_i - \widetilde Q_i = \frac{\alpha_s}{4\pi} \sum_r \tilde \omega_r Q_r.
\end{equation}
That is the EVOs are defined by the shifts between the 1-loop insertions of an operator and 
its Fierzed conjugate. Finally, to obtain the 1-loop basis relations, 
one have to identify the Fierz transformations used in each tree-level relation 
and include the ${\rm EV_i}$ wherever the Fierz relations have been employed. 
Hence, at 1-loop level the basis change relation becomes
\begin{equation}
O_{JMS} = \hat R Q_{BMU} \,,
\end{equation}
with 
\begin{equation} \label{eq:1looptras}
\hat R = \hat R^{(0)} + \frac{\alpha_s}{4\pi} \hat R^{(1)}.
\end{equation}
The transformation rules at 1-loop level can be summarized as
\begin{enumerate}
\item If no Fierz transformations are required at the tree-level, 
then the corresponding blocks in $\hat R^{(0)}$ and  $\hat R$ and are simply 
equal.
\item If at tree-level the Fierz transformations are required but the corresponding
shifts vanish, then the corresponding blocks in $\hat R^{(0)}$ and  $\hat R$ are still the same. 
\item  In certain blocks the necessity of performing Fierz transformation would
introduce the shifts which will introduce EVOs and hence contribute to
the $\hat R^{(1)}$.
\end{enumerate}
\section{JMS to BMU translation at 1-loop}
\begin{figure*}
\begin{center}
\includegraphics[width=11cm, height=4cm]{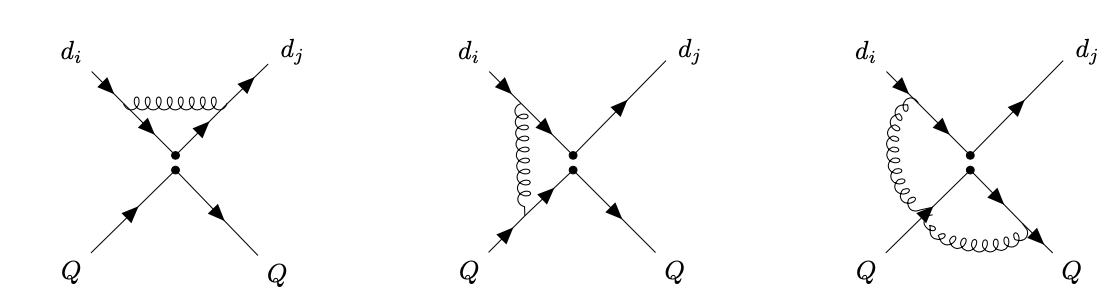}  \label{fig:loop1}
\\[0.5cm]
\includegraphics[width=6cm, height=4cm]{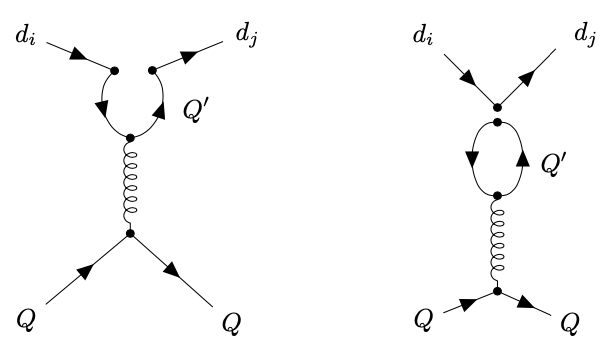}  \label{fig:loop2}
\caption{Current-current (upper panel) and penguin insertions (lower panel).}
\end{center}
\end{figure*}
Now we present a concrete example of transformation between the BMU and JMS bases which have been used to
compute the 2-loop anomalous dimensions and SMEFT to WET matching at the 1-loop level, respectively. 
For simplicity, we focus on the operators involved in the $\Delta F=1$ flavour transitions. The method however can also
be applied for the other kind operators in the WET \cite{Aebischer:2020dsw}. The relevant operators for $\Delta F=1$ transitions in the 
BMU basis are listed as follows. 
The well known SM operators $Q_1-Q_{10}$ in the BMU basis are give by

\begin{equation}
\begin{aligned}
  \OpL{1} &
  = \OpL[\text{VLL},u]{1}
  = (\bar d_j^\alpha \gamma_\mu P_L u^\beta)
    (\bar u^\beta    \gamma^\mu P_L d_i^\alpha) ,
\\
  \OpL{2} &
  = \OpL[\text{VLL},u]{2}
  = (\bar d_j^\alpha \gamma_\mu P_L u^\alpha)
    (\bar u^\beta    \gamma^\mu P_L d_i^\beta) ,
\\
  \OpL{3} & = (\bar d_j^\alpha \gamma_\mu P_L d_i^\alpha)
    \!\sum_{q}(\bar q^\beta    \gamma^\mu P_L \, q^\beta) , 
\\
  \OpL{4} & = (\bar d_j^\alpha \gamma_\mu P_L d_i^\beta)
    \!\sum_{q}(\bar q^\beta    \gamma^\mu P_L \, q^\alpha) ,
\\
  \OpL{5} & = (\bar d_j^\alpha \gamma_\mu P_L d_i^\alpha)
    \!\sum_{q}(\bar q^\beta    \gamma^\mu P_R \, q^\beta) ,
\\
  \OpL{6} & = (\bar d_j^\alpha \gamma_\mu P_L d_i^\beta)
    \!\sum_{q}(\bar q^\beta    \gamma^\mu P_R \, q^\alpha) ,
\end{aligned}
\end{equation}

\begin{equation}
  \label{eq:QED-peng-op}
\begin{aligned}
  \OpL{7} & = \frac{3}{2}\,(\bar d_j^\alpha \gamma_\mu P_L d_i^\alpha)
      \!\sum_{q} \! Q_q \, (\bar q^\beta \gamma^\mu P_R \, q^\beta) ,  
\\
  \OpL{8} & = \frac{3}{2}\,(\bar d_j^\alpha \gamma_\mu P_L d_i^\beta)
      \!\sum_{q} \! Q_q \, (\bar q^\beta \gamma^\mu P_R \, q^\alpha) ,
\\
  \OpL{9} & = \frac{3}{2}\,(\bar d_j^\alpha \gamma_\mu P_L d_i^\alpha)
      \!\sum_{q} \! Q_q \, (\bar q^\beta \gamma^\mu P_L \, q^\beta) ,
\\
  \OpL{10} & =\frac{3}{2}\,(\bar d_j^\alpha \gamma_\mu P_L d_i^\beta)
      \!\sum_{q} \! Q_q \, (\bar q^\beta \gamma^\mu P_L \, q^\alpha) .
\end{aligned}
\end{equation}
In general, for the beyond the SM physics, there are in total 40 $\Delta F=1$ operators plus their chirality 
flipped partners. The complete list of BMU operators including the BSM ones can be found in Ref.~\cite{Aebischer:2022tvz, Aebischer:2021raf}.
The JMS operators are given in Table~\ref{tab:jms}.
\begin{table*}[htb]
\centering
\begin{tabular}{||c|c||c|c||}
\hline\hline
\multicolumn{2}{||c||}{$(\bar LL)(\bar LL)$} & \multicolumn{2}{|c||}{$(\bar RR)(\bar RR)$}  \\ \hline
$\opL[V,LL]{dd}{prst}$ & $(\bar{d}_L^p \gamma_\mu d_L^r) (\bar{d}_L^s \gamma^\mu d_L^t)$ &
$\opL[V,RR]{dd}{prst}$& $(\bar{d}_R^p \gamma_\mu d_R^r) (\bar{d}_R^s \gamma^\mu d_R^t)$  \\
$\opL[V1,LL]{ud}{prst}$ & $(\bar{u}_L^p \gamma_\mu u_L^r) (\bar{d}_L^s \gamma^\mu d_L^t)$ &
$\opL[V1,RR]{ud}{prst}$ & $(\bar{u}_R^p \gamma_\mu u_R^r) (\bar{d}_R^s \gamma^\mu d_R^t)$  \\
$\opL[V8,LL]{ud}{prst}$& $(\bar{u}_L^p \gamma_\mu T^A u_L^r) (\bar{d}_L^s \gamma^\mu T^A d_L^t)$  &
$\opL[V8,RR]{ud}{prst}$& $(\bar{u}_R^p \gamma_\mu T^A u_R^r) (\bar{d}_R^s \gamma^\mu T^A d_R^t)$ \\  \hline\hline
\multicolumn{2}{||c||}{$(\bar LL)(\bar RR)$} & \multicolumn{2}{|c||}{$(\bar LR)(\bar LR)$+ \text{h.c.}}  \\ \hline
$\opL[V1,LR]{dd}{prst}$&$(\bar{d}_L^p \gamma_\mu d_L^r) (\bar{d}_R^s \gamma^\mu d_R^t)$&
$\opL[S1,RR]{dd}{prst}$& $(\bar{d}_L^p d_R^r) (\bar{d}_L^s d_R^t)$\\
$\opL[V8,LR]{dd}{prst} $&$(\bar{d}_L^p \gamma_\mu T^A d_L^r) (\bar{d}_R^s \gamma^\mu T^A d_R^t)$&
$\opL[S8,RR]{dd}{prst}$& $(\bar{d}_L^p T^A d_R^r) (\bar{d}_L^s T^A d_R^t)$\\
$\opL[V1,LR]{ud}{prst}$&$ (\bar{u}_L^p \gamma_\mu u_L^r) (\bar{d}_R^s \gamma^\mu d_R^t)$&
$\opL[S1,RR]{ud}{prst}$& $(\bar{u}_L^p u_R^r) (\bar{d}_L^s d_R^t)$\\
$\opL[V8,LR]{ud}{prst}$&$(\bar{u}_L^p \gamma_\mu T^A u_L^r) (\bar{d}_R^s \gamma^\mu T^A d_R^t) $&
$\opL[S8,RR]{ud}{prst}$& $(\bar{u}_L^p T^A u_R^r) (\bar{d}_L^s T^A d_R^t)$\\
$\opL[V1,LR]{du}{prst}$&$(\bar{d}_L^p \gamma_\mu d_L^r) (\bar{u}_R^s \gamma^\mu u_R^t)$&
$\opL[S1,RR]{uddu}{prst}$& $(\bar{u}_L^p d_R^r) (\bar{d}_L^s u_R^t)$\\
$\opL[V8,LR]{du}{prst}$&$(\bar{d}_L^p \gamma_\mu T^A d_L^r) (\bar{u}_R^s \gamma^\mu T^A u_R^t)$&
$\opL[S8,RR]{uddu}{prst}$& $(\bar{u}_L^p T^A d_R^r) (\bar{d}_L^s T^A u_R^t)$\\
$\opL[V1,LR]{uddu}{prst}$&$(\bar{u}_L^p \gamma_\mu d_L^r) (\bar{d}_R^s \gamma^\mu u_R^t)$ + h.c.&
& \\
$\opL[V8,LR]{uddu}{prst}$&$(\bar{u}_L^p \gamma_\mu T^A d_L^r) (\bar{d}_R^s \gamma^\mu T^A u_R^t)$ + h.c.&
& \\
\hline \hline
\end{tabular}
\caption{\small Non-leptonic $\Delta F=1$ operators (baryon and lepton number conserving) 
in the JMS basis \cite{Jenkins:2017jig}. } \label{tab:jms}
\end{table*}
As explained in the previous section, first one have to obtain the EV shifts in one of the basis under consideration. 
In this example we work out the shifts for the BMU operators
\be
\label{eq:ruleQ1Q2}
  \OpL{1} = \OpLt{1}, \qquad
  \OpL{2} = \OpLt{2} + \frac{1}{3}\frac{\alS}{4\pi} P,
\ee
with
\be
  P = \OpL{4} + \OpL{6} - \frac{1}{3}(\OpL{3} + \OpL{5}).
\ee
\be\label{eq:ruleQ3Q6}
  \OpL{3} = \OpLt{3}+ \frac{2}{3}\frac{\alS}{4\pi} P    , \qquad
\OpL{4} = \OpLt{4} - \frac{N_f}{3}\frac{\alS}{4\pi} P, 
\ee
\be
\OpL{5} = \OpLt{5}, \qquad  \OpL{6} = \OpLt{6}\,,
\ee
\be\label{eq:ruleQ7Q10}
 \OpL{7} = \OpLt{7}, \qquad  \OpL{8} = \OpLt{8}, 
\ee
\be
  \OpL{9} = \OpLt{9}- \frac{1}{3}\frac{\alS}{4\pi} P    , \qquad
  \OpL{10} = \OpLt{10} - \frac{1}{3}(N_u-\frac{N_d}{2})\frac{\alS}{4\pi} P.
  \ee
These shifts have been obtained in the NDR-MSbar scheme as defined in Buras+Weisz \cite{Buras:1989xd} that uses Greek method.
We find that the Fierz transformations on the VLR operators $\OpL{k}$ with $k=5-8$ do not bring any 
contributions from evanescent operators. In the case of the BSM operators $\OpL{k}$ with $k=11-18$ only the Fierz
transformation on  $\OpL{11}$ brings a contribution from evanescent operators so that
\be \label{eq:ruleQ11}
\OpL{11} = \OpLt{11} +\frac{2}{3}\frac{\alS}{4\pi} P,
\qquad \OpL{k} = \OpLt{k},\quad k=12-18\,.
\ee
In the same way one can obtain the shifts for the current-current insertions, see 
for example Eqs. (28)-(31) of Ref.~\cite{Aebischer:2022tvz}. Having these shifts it is straightforward to obtain 
the 1-loop transformation matrix defined in Eq.~\eqref{eq:1looptras}.
\section{Summary}
In a given EFT, it might be useful to work simultaneously with more than one operator basis.
For instance, the JMS basis for WET is useful for the matching to SMEFT, however, for the running below the EW scale at the 
NLO level, the BMU is more convenient. We have presented a simple procedure to perform the transformations of the bases at
the 1-loop level. Formal techniques are also known in the literature\cite{Dugan:1990df, Buras:1989xd, Aebischer:2021raf}.
\begin{acknowledgments}
I am thankful to Andrzej J. Buras and Jason Aebischer for a pleasant collaboration. I am financially supported by the Alexander von Humboldt Foundation’s postdoctoral
research fellowship.
\end{acknowledgments}

\bigskip 

\end{document}